\documentclass[usenatbib,usegraphicx]{mn2e}
\usepackage[colorlinks,citecolor=blue,linkcolor=blue,urlcolor=blue,dvips]{hyperref}
\usepackage{breakurl}
\usepackage{amssymb}
\usepackage{amsmath}
\usepackage{mathrsfs}

\newcommand{\apj}   {ApJ}
\newcommand{\apjl}  {ApJ}
\newcommand{\apjs}  {ApJS}

\newcommand{\mnras} {MNRAS}

\newcommand {\B} {\begin{equation}}
\newcommand {\E} {\end{equation}}
\newcommand {\bea} {\begin{eqnarray}}
\newcommand {\eea} {\end{eqnarray}}

\newcommand{\like}{\mathscr{L}}

\defcitealias{Nalewajko2013}{Paper I}

\graphicspath{{figures/}}

\pagerange{\pageref{firstpage}--\pageref{lastpage}} \pubyear{2014}

\title[Turbulent spectra of blazars]{Turbulent spectra of the brightest gamma-ray flares of blazars}

\author[Kohler \& Nalewajko]{Susanna~Kohler$^1$\thanks{\tt{e-mail: kohlers@colorado.edu}} and Krzysztof~Nalewajko$^{1,2,3}$\\
$^1$JILA, University of Colorado and National Institute of Standards and Technology, 440 UCB, Boulder, CO 80309, USA\\
$^2$Kavli Institute for Particle Astrophysics and Cosmology, Stanford University and SLAC National Accelerator Laboratory,\\
2575 Sand Hill Road M/S 29, Menlo Park, CA 94025, USA\\
$^3$NASA Einstein Postdoctoral Fellow
}

\begin{document}
\label{firstpage}

\maketitle

\begin{abstract}
We investigate the spectral properties of the brightest gamma-ray flares of blazars detected by the \emph{Fermi} Large Area Telescope. We search for the presence of spectral breaks and measure the spectral curvature on typical time scales of a few days. We identify significant spectral breaks in fewer than half of the analyzed flares, but their parameters do not show any discernible regularities, and in particular there is no indication for gamma-ray absorption at any fixed source-frame photon energy. More interestingly, we find that the studied blazars are characterized by significant spectral variability. Gamma-ray flares of short duration are often characterized by strong spectral curvature, with the spectral peak located above 100 MeV. Since these spectral variations are observed despite excellent photon statistics, they must reflect temporal fluctuations in the energy distributions of the emitting particles. We suggest that highly regular gamma-ray spectra of blazars integrated over long time scales emerge from a superposition of many short-lived irregular components with relatively narrow spectra. This would imply that the emitting particles are accelerated in strongly turbulent environments.
\end{abstract}


\section{Introduction}
\label{intro}

Blazars, a class of active galactic nuclei, belong to the brightest cosmic sources of high-energy ($\sim{\rm GeV}$) gamma-ray radiation. Their gamma-ray emission is produced by non-thermal populations of highly energetic particles in relativistic jets. The mechanism of particle acceleration responsible for the emergence of those populations is not yet understood, but detailed analysis of the gamma-ray data on blazars has the potential to provide more constraints on these processes.

The \emph{Fermi} Large Area Telescope (LAT) has unique capabilities in high-energy astronomy: a very broad spectral range ($\sim 100\;{\rm MeV} - 100\;{\rm GeV}$), a very wide field of view ($\sim 60^\circ$), and the ability to scan the entire sky every 3 hours \citep{Atwood2009}. After several years of its mission, it has collected a vast amount of data on the temporal and spectral behaviour of blazars. Of particular importance are data on the brightest blazars, known as Flat-Spectrum Radio Quasars (FSRQs). Their broad-band spectral energy distributions (SEDs) are strongly dominated by the MeV-GeV gamma-ray band \citep{Fossati1998}, and they reach gamma-ray fluxes of $10^{-5}\;{\rm ph\,s^{-1}\,cm^{-2}}$ and higher \citep{Abdo2011}. At such high fluxes, they can be analyzed at very high temporal and/or spectral resolution.

The gamma-ray spectra of FSRQs are typically steep, with average photon indices of $\Gamma \simeq 2.2 - 2.7$ \citep{Abdo2010a,Ackermann2011}. However, these spectra are not consistent with simple power laws (SPLs); they can be better fit by either a log parabola (LP) model, or a broken power law (BPL) model. In many blazars, spectral breaks are routinely identified, and they can be parametrized by the observed photon energy at the break $E_{\rm br,obs}$ and the change in the photon index $\Delta\Gamma$. In the initial \emph{Fermi}/LAT data on 3C~454.3, the brightest blazar of the \emph{Fermi} era, a break was identified in spectra integrated over a month of observations with $E_{\rm br,obs} \simeq 2.4\;{\rm GeV}$ and $\Delta\Gamma \simeq 1.2$ \citep{Abdo2009}. In the subsequent studies of 3C~454.3, breaks were found with $E_{\rm br,obs} \simeq 1.0 - 2.8\;{\rm GeV}$ and $\Delta\Gamma \simeq 0.6 - 1.0$ \citep{Ackermann2010,Abdo2011}. In other blazars, spectral breaks were identified with break energies $E_{\rm br} \simeq 1.6 - 10\;{\rm GeV}$ in the source frame \citep{Abdo2010a}.

It was immediately recognized that such sharp spectral breaks cannot be due to a transition to the fast-cooling regime of electron energy distribution, which predicts a change in photon index of $\Delta\Gamma = 0.5$ \citep{Abdo2009}. For a power-law distribution of electrons, it was demonstrated that transition to the Klein-Nishina regime of inverse Compton scattering does not produce a sharp spectral break, but rather a smooth cut-off \citep{Tave08,Ackermann2010}. However, considering a curved electron energy distribution, a spectral break can be obtained in this regime \citep{Cerruti2013}. The conclusion of \cite{Abdo2009} was that these breaks are most likely due to a break in the underlying electron energy distribution. However, \cite{Finke2010} noted that this scenario is incompatible with the observed optical/UV spectra of 3C~454.3. Instead, they proposed that the observed breaks could be explained by a superposition of two spectral components, produced by Comptonization of broad emission lines and direct radiation of the accretion disk.

An interesting alternative was proposed by \cite{Poutanen2010}, who argued that the gamma-ray spectral breaks in blazars could arise from absorption by recombination continua of ionized helium (He II). This model predicts that the observed breaks should be close to $E_{\rm br} \simeq 5\;{\rm GeV}$ in the source frame. In a subsequent study of the gamma-ray spectra of 3C~454.3 at several different flux levels, \cite{Stern2011} found a weak anti-correlation between the optical depth for He II continuum, inferred from $\Delta\Gamma$, and the total gamma-ray flux. The implication of strong absorption by ionized helium is that gamma-ray emitting regions should be located very close ($\sim 0.1\;{\rm pc}$) to the central black hole, possibly shifting to larger distances with increasing gamma-ray flux.

More recent studies have cast doubt on the universal value of break energies measured in the source frame \citep{Harris2012} and have also suggested that some breaks identified in early studies using \emph{Fermi}/LAT data could be artifacts of inaccurate instrument response functions \citep{Harris2014}. \cite{Stern2014} relaxed their original claim after updated analysis and now argue that only two FSRQs of the nine that they studied display spectral breaks at a consistent energy $E_{\rm br} \simeq 5$ GeV, while most of these sources show breaks at $E_{\rm br} \simeq 20$ GeV interpreted as due to absorption by hydrogen recombination continua.

Another interesting aspect of the gamma-ray spectra of blazars is the spectral curvature, which can be probed by fitting log-parabola models to the spectra. Such models return two parameters --- the curvature index $\beta$, and the peak photon energy $E_{\rm peak}$ (see definitions in \S \ref{sec_data}). 
Typical spectra of FSRQs have $E_{\rm peak} < 100\;{\rm MeV}$ with $\beta \sim 0.05 - 0.3$ \citep{Ackermann2011,Harris2014}.
It is generally thought that the gamma-ray spectra of blazars are stable, but this problem was not investigated very thoroughly.
\cite{Stern2011} showed for 3C~454.3 that the curvature index is roughly independent of the gamma-ray flux, but that the peak energy is strongly correlated with the gamma-ray flux, falling in the range of $\sim 10 - 100\;{\rm MeV}$.

In \cite{Nalewajko2013} (hereafter \citetalias{Nalewajko2013}), a sample was selected of the brightest gamma-ray flares of blazars during the first four years of the \emph{Fermi} mission. The study in Paper I focused on the temporal properties of the flares, such as duration and time asymmetry; however, time variations of the gamma-ray photon index were also investigated. It was demonstrated that many blazar flares exhibit significant variations of the photon index, and that some very short flares (of duration $\lesssim 1\;{\rm d}$) show relatively hard spectra, with $\Gamma \lesssim 2$. The most notable example of such flares is the MJD~55854 event in PKS~1510--089 \citep{Saito2013}, also included in our sample. Additional similar events have been identified recently, and they may represent an important new class of gamma-ray events observed in luminous blazars.

Here, we analyze the spectral properties of the sample of flares selected in \citetalias{Nalewajko2013}. Rather than attempting to provide a general model of the observed gamma-ray spectra, we seek to identify and analyze features that are consistent with any of the three standard spectral models (SPL, BPL or LP). In particular, we focus on two aspects: 1) the occurrence and properties of spectral breaks, and 2) spectral curvature and spectral variations in the sample. We begin by describing our analysis of the spectral properties of the flares in Section \ref{sec_data}. We then present our results on the spectral breaks and the spectral curvature in Section \ref{sec_res}. This is followed by a discussion in Section \ref{sec_dis}, and conclusions in Section \ref{sec_con}.

\section{Data analysis}
\label{sec_data}

We investigate the sample of 40 bright blazar flares selected in \citetalias{Nalewajko2013} (details of the selection procedure are given therein). Each flare is a period of time when the observed photon flux exceeds half of the peak flux. The minimum peak flux for the sample is $7.1\times 10^{-6}\;{\rm ph\,s^{-1}\,cm^{-2}}$. These flares have durations $0.5-10$ days, and they are produced by only five blazars: 3C~454.3, PKS~1510--089, PKS~1222+216, 3C~273, and PKS~0402--362.

\begin{figure*}
\center
\includegraphics[width=\textwidth]{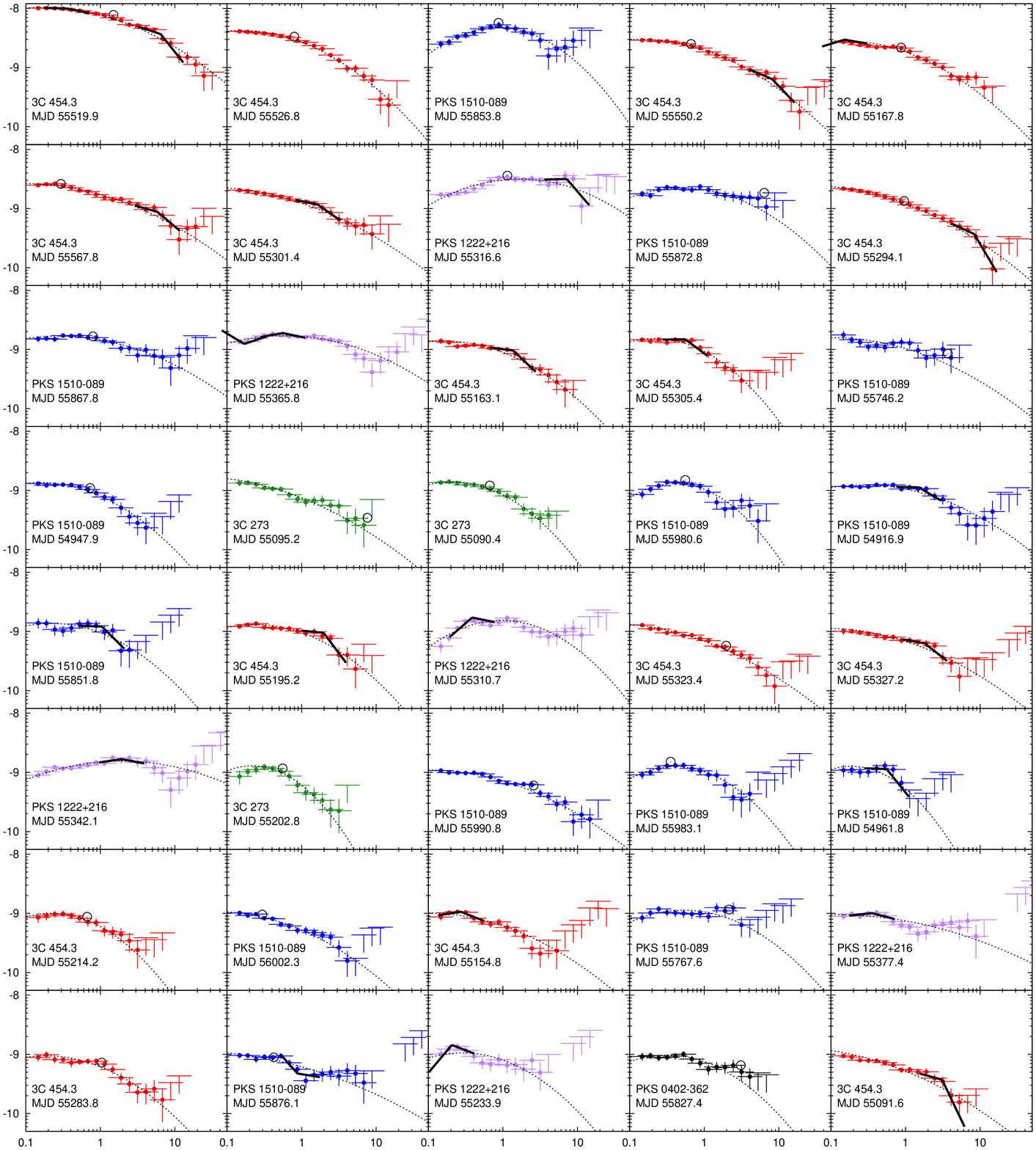}
\caption{Time-integrated flare spectra displaying the $\nu F_\nu$ flux (${\rm erg \,s^{-1}\,cm^{-2}}$) vs observed energy (GeV) of the 40 brightest gamma-ray blazar flares detected by \emph{Fermi}/LAT. Statistically significant primary spectral breaks (those identified with a BPL fit to the entire energy range $0.1 - 10\;{\rm GeV}$) and secondary spectral breaks (those identified with a BPL fit to only the energy range above or below the primary break energy) are indicated by broken lines that illustrate the mean value of the spectral index $\Gamma$ on either side of the break. Primary breaks that are not statistically significant are indicated by open circles. Dotted lines show the best-fit primary log-parabola models. The color of the spectrum indicates the host blazar.}
\label{fig:top40}
\end{figure*}

For each flare period, we integrate the binned SED using the standard analysis tool {\tt gtlike} from the \emph{Fermi}/LAT {\tt ScienceTools} software package ({\tt v9r27p1}). In the analysis, we use the instrument response function {\tt P7SOURCE\_V6}, the Galactic diffuse emission model {\tt gal\_2yearp7v6\_v0}, the isotropic background model {\tt iso\_p7v6source}, events of the {\tt SOURCE} class, the region of interest of radius $10^\circ$, and all background sources from the 2FGL catalog \citep{Nolan2012} within $15^\circ$. The energy bins are of equal width and uniformly distributed on a logarithmic scale, and they overlap with the logarithmic shift equal to 1/3 of the logarithmic bin length. In cases of insufficient photon statistics, standard $2\sigma$ upper limits are calculated. The results are shown in Fig \ref{fig:top40} in order of decreasing peak flux (as in \citetalias{Nalewajko2013}).

Next, for each flare we performed maximum likelihood analysis on the unbinned data in the energy range between $E_{\rm min} = 0.1\;{\rm GeV}$ and $E_{\rm max} = 10\;{\rm GeV}$ by fitting spectral models of an SPL, BPL, and LP. The maximum energy limit $E_{\rm max}$ is chosen due to relatively short integration time scales, on which only a handful of $>10$ GeV photons are detected for each flare.

For the SPL fit, we used the PowerLaw2 spectral model from the \emph{Fermi} analysis tools,\footnote{\url{http://fermi.gsfc.nasa.gov/ssc/data/analysis/scitools/source_models.html}}
	\B
		 \frac{\text{d}N}{\text{d}E} = \frac{N (1-\Gamma) E^{-\Gamma}}{E_{\rm max}^{1-\Gamma}-E_{\rm min}^{1-\Gamma}},
	\E
where $\text{d}N/\text{d}E$ is the differential photon flux as a function of photon energy, $\Gamma$ is the spectral index, and $E_{\rm min}$ and $E_{\rm max}$ are the fixed lower and upper bounds of the energy range. Using this model rather than the basic power law fit allowed us to avoid artificially selecting a normalization energy and instead fix the energy range over which we wished to apply the model. In similar fashion, we used the BrokenPowerLaw2 spectral model for the BPL fit,
	\B
		 \frac{\text{d}N}{\text{d}E} = N_0 \times 
	 	 \begin{cases}
	 		 (E/E_{\rm br})^{-\Gamma_1} & \text{if } E < E_{\rm br} \\
 			 (E/E_{\rm br})^{-\Gamma_2}      & \text{otherwise},
	 	 \end{cases}
	\E
where $N_0$ is a normalization factor, $E_{\rm br}$ is the break energy, and $\Gamma_1$ and $\Gamma_2$ are the spectral indices in the range where $E<E_{\rm br}$ and $E>E_{\rm br}$ respectively. For the LP fit, we used the LogParabola model,
	\B
	 \frac{\text{d}N}{\text{d}E} = N_0 \left( \frac{E}{E_0} \right) ^{-(\alpha + \beta \text{log}(E/E_0))},
	\E
where $E_0$ is a fixed pivot energy, $N_0$ is the normalization, $\beta$ describes the spectral curvature, and $\alpha$ is a parameter related to the peak energy,
	\B
	E_{\rm peak} = E_0  \exp \left( \frac{2-\alpha}{2\beta} \right).
	\E
We set the pivot energy to $E_0= 500$ MeV, in keeping with other analyses (e.g. \cite{Harris2012}), after first testing that moving this pivot doesn't have a significant impact on the likelihood value for the fit.

In analyzing the fit of these three models to the data, we did not seek to determine whether the models provided a good global description of the observed spectra. Our goal was instead to determine which of the three models provided the \emph{best} fit to the data. To do this, we performed an Akaike Information Criterion (AIC) test, determining the AIC value for each model:
	\B
	\text{AIC} = 2k - 2 \text{ln} \like,
	\E
where $k$ is the number of free parameters in the model, and $\like$ is the maximized value of the likelihood function for the estimated model. The preferred model is then the one with the minimum AIC, and if the difference in the AIC between two fits is such that $\Delta \text{AIC}>2$, the preference for the model with the minimum AIC is generally considered to be statistically significant (refer to \cite{Harris2014}, \cite{Bozdogan1987}, and \cite{Lewis2011} for more information on the AIC and $\Delta$AIC). In our case, if the BPL fit had a $\Delta \text{AIC}>2$ relative to the other fits, we declared this spectral break significant and recorded the fit parameters in Table \ref{tab:sigbreaks}.

Uncertainties in most fit parameters were obtained directly from the fits, with the exception of the parameter $E_{\rm br}$. Uncertainties in $E_{\rm br}$ were instead estimated from the statistical uncertainty corresponding to $-2 \Delta L = 1$ for the 1-$\sigma$ confidence region, where $L$ is the log-likelihood function (see \cite{Ackermann2010}, \cite{Rolke2005} for discussion of this technique). In cases where $E_{\rm br}$ was unbound, we rejected the break.

In many cases, the best-fit BPL model with break energy $E_{\rm br1}$ did not yield a significant spectral break, while a candidate break could be seen at $E_{\rm br2} \ne E_{\rm br1}$. A good example of this is flare \#10 (3C 454.3, MJD 55294) with $E_{\rm br1} \sim 0.9\;{\rm GeV}$ and $E_{\rm br2} \sim 9\;{\rm GeV}$. A candidate break at $E_{\rm br2}$ may not be identified by the global BPL fit due to the overall spectral curvature. This was noted by \cite{Stern2011}, who found that spectral breaks can be identified more robustly when the underlying background spectral model is a log-parabola rather than a power-law. In order to account for spectral breaks that could have been missed by the global (\emph{primary}) BPL fits, for every flare in the sample we performed an additional search for \emph{secondary spectral breaks} by fitting the BPL, SPL and LP models separately in the energy ranges $E_{\rm min} < E < E_{\rm br1}$ (\emph{low-energy secondary}) and $E_{\rm br1} < E < E_{\rm max}$ (\emph{high-energy secondary}). In the example of flare \#10, a secondary break was indeed identified at $E_{\rm br2} \sim 8.4\;{\rm GeV}$. All statistically significant secondary spectral breaks are included in Table \ref{tab:sigbreaks}.

\section{Results}
\label{sec_res}

\subsection{Spectral breaks}
\label{sec_res_breaks}

In analyzing the spectra of the 40 flares, we found that none were best fit by a SPL model, 31 were best fit by a BPL model, and nine were best fit by a LP model. Of the 31 BPL-favored spectra, the BPL model was significantly favored over other models in 15 cases. The breaks from two of these cases were rejected because they were unbound, but the 13 remaining primary breaks are recorded in Table \ref{tab:sigbreaks}. Ten additional statistically significant breaks were found by the secondary analysis. They are also recorded in the table (labeled to identify them as secondary breaks found either on the low- or high-energy side of the primary break of the spectrum), for a total of 23 significant breaks detected in the 40 flares. Figure \ref{fig:top40} displays the binned spectrum for each of the 40 flares and illustrates the location of any significant primary and secondary breaks, as well as the values of the spectral index $\Gamma$ on either side of the breaks. The primary breaks that are not significant are also noted on the spectra for reference. It is important to note that the models were only fit to the data below 10 GeV; in cases where spectral breaks are evident, the observed structure of the spectra could extend into the very high energy (VHE) range, but this cannot be determined due to very low photon statistics above 10 GeV.

\begin{figure}
\center
\includegraphics[width=0.9\columnwidth]{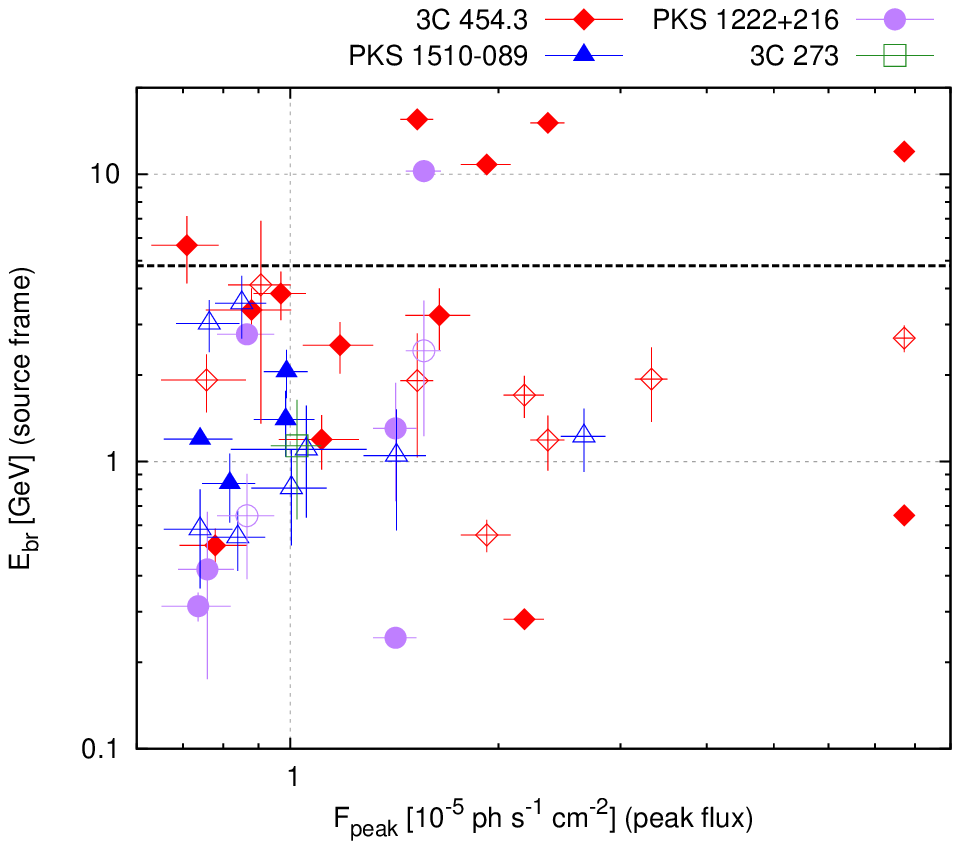}
\includegraphics[width=0.9\columnwidth]{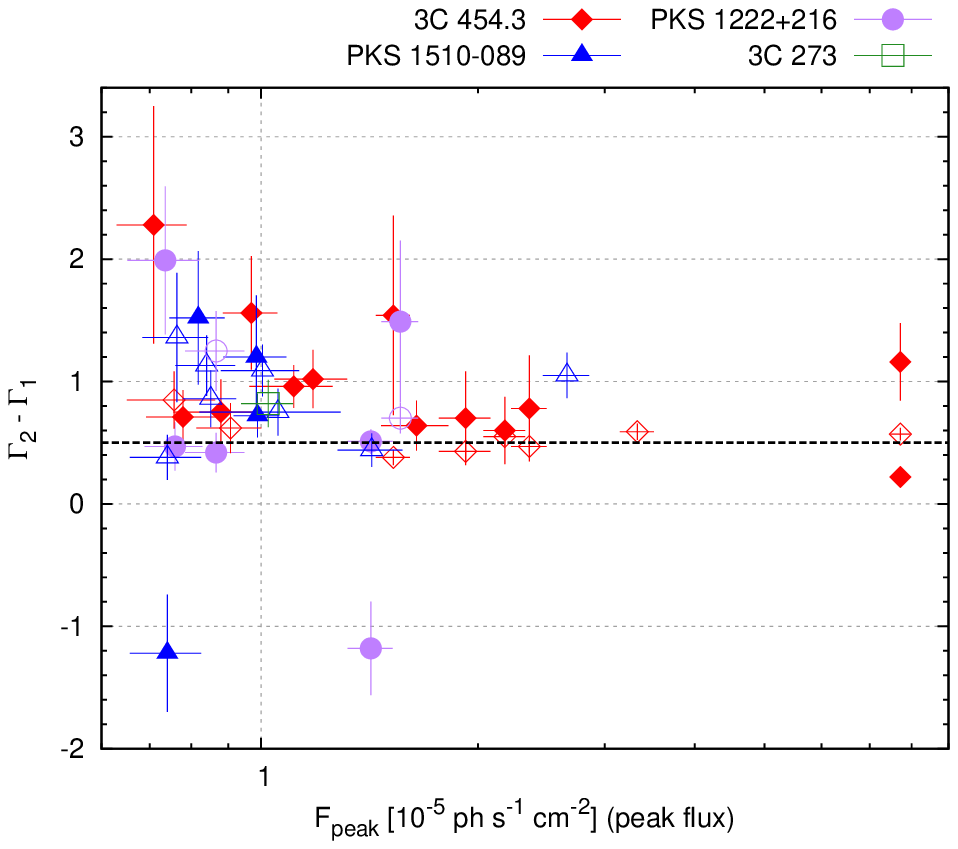}
\includegraphics[width=0.9\columnwidth]{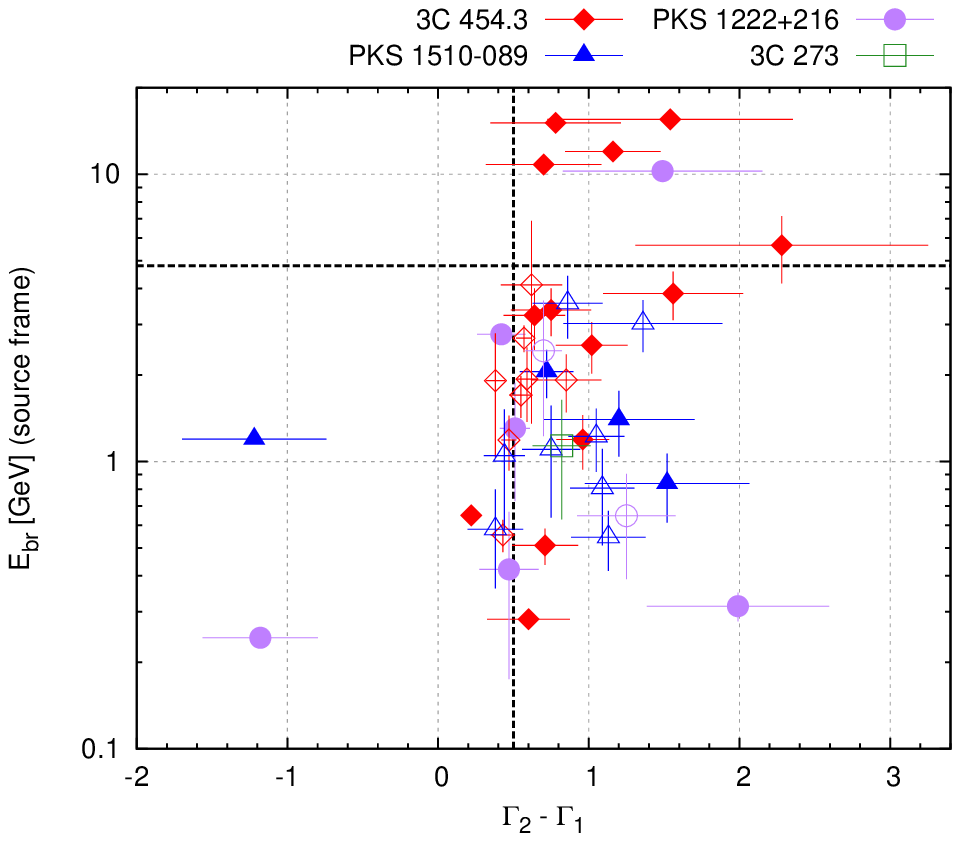}
\caption{Distributions of the source-frame break energy $E_{\rm br}$ (GeV), the change in photon index $\Delta \Gamma = \Gamma_2 - \Gamma_1$, and the peak photon flux $F_{peak}$ ($10^{-5}\;{\rm ph\,s^{-1}\,cm^{-2}}$), plotted for all identified spectral breaks in the top 40 \emph{Fermi}/LAT gamma-ray flares. The shape and color of each point indicate the host blazar. Solid symbols indicate statistically significant primary and secondary breaks, whereas unfilled symbols indicate primary breaks that were not statistically significant.}
\label{fig:ebr}
\end{figure}

As part of the analysis of the spectral breaks of these flares, we sought to test the `double absorber' model put forward by \cite{Poutanen2010}. This model predicts two increases in the opacity of the broad-line region to gamma-ray-energy photons: one at $\sim 5$ GeV (in the source frame) due to He II recombination, and one at $\sim 20$ GeV due to H I recombination. These opacity increases result in changes in the photon index that should be seen clearly as breaks in the flare spectra. Because our analysis does not extend beyond 10 GeV due to relatively short integration time scales, we do not attempt to comment on the presence of the proposed break due to H I, however, we analyze the breaks we found in the $0.1 -10$ GeV range to search for a preference for breaks near 5 GeV. In Figure \ref{fig:ebr}, we plot the distribution of the redshift-corrected break energies against the peak flux of the flare. As can be seen, there is no indication for a break preference at 5 GeV or any other fixed energy in the source frame. Instead, the break energies seem to be spread uniformly between $0.2\;{\rm GeV} < E_{\rm br} < 20\;{\rm GeV}$. We do not find any indication for a correlation between $E_{\rm br}$ and other parameters of the flares, such as peak flux, duration, or time asymmetry.

In Figure \ref{fig:ebr}, we also plot the distribution of the break amount, or change in the photon index $\Delta\Gamma$ at the break, vs. peak flux. For completeness, we also plot the distribution of $\Delta\Gamma$ vs. $E_{\rm br}$. For two spectral breaks, we found $\Delta\Gamma < 0$ (negative breaks), indicating spectral hardening with increasing energy. Both of these breaks are secondary, and they reflect local spectral features. For the positive breaks, we find that $0.5 \lesssim \Delta\Gamma \lesssim 2$. The value of $\Delta\Gamma = 0.5$ can be understood theoretically as resulting from a break in the electron energy distribution $N(\gamma) \propto \gamma^{-p}$ by $\Delta p = 1$, which could be due to the transition from inefficient to efficient cooling \citep{1993ApJ...416..458D}. While a substantial number of significant spectral breaks are close to this value in our data, roughly the same number of breaks are clearly inconsistent with it. The value of $\Delta\Gamma$ does not seem to be correlated with other parameters of the flare such as observed flux, duration or time asymmetry. \cite{Stern2011} showed that in the case of 3C~454.3 the value of $\Delta\Gamma$, interpreted as a measure of optical depth for absorption from the He II continuum, is weakly anticorrelated with the gamma-ray luminosity (or flux). We do not find any evidence for this in our results.

\subsection{Spectral curvature}
\label{sec_res_curv}

In analyzing the flare spectra, our second goal was to examine the spectral curvature of all 40 flares, regardless of which model provided the best fit to the data. Table \ref{tab:logpar} contains flare details and fit parameters from the LP model applied to each of the 40 flares across the full energy range. The spectral curvature parameter $\beta$ spans the range of $\sim 0.05 - 0.3$, which is consistent with the finding that none of the flares has a gamma-ray spectrum best described by SPL model. It also means that all the gamma-ray spectra are concave, which is natural for spectra dominated by a single spectral component. Interestingly, we find a very broad range of the spectral peak values, with $0.03\;{\rm GeV} < E_{\rm peak} < 2\;{\rm GeV}$. The very low values $E_{\rm peak} < 0.1\;{\rm GeV}$ are often insignificant, as they are derived from an extrapolation of the LP model beyond the observed energy range. Nevertheless, these low estimates correspond to the typical shape of the gamma-ray spectra of FSRQ blazars, where the unobserved SED peak is generally thought to lie somewhere in the $\sim 1-10\;{\rm MeV}$ range \citep{Fossati1998,Abdo2010b}. The cases where $E_{\rm peak} > 0.1\;{\rm GeV}$ would in general be considered to be atypical.

\begin{figure}
\center
\includegraphics[width=0.9\columnwidth]{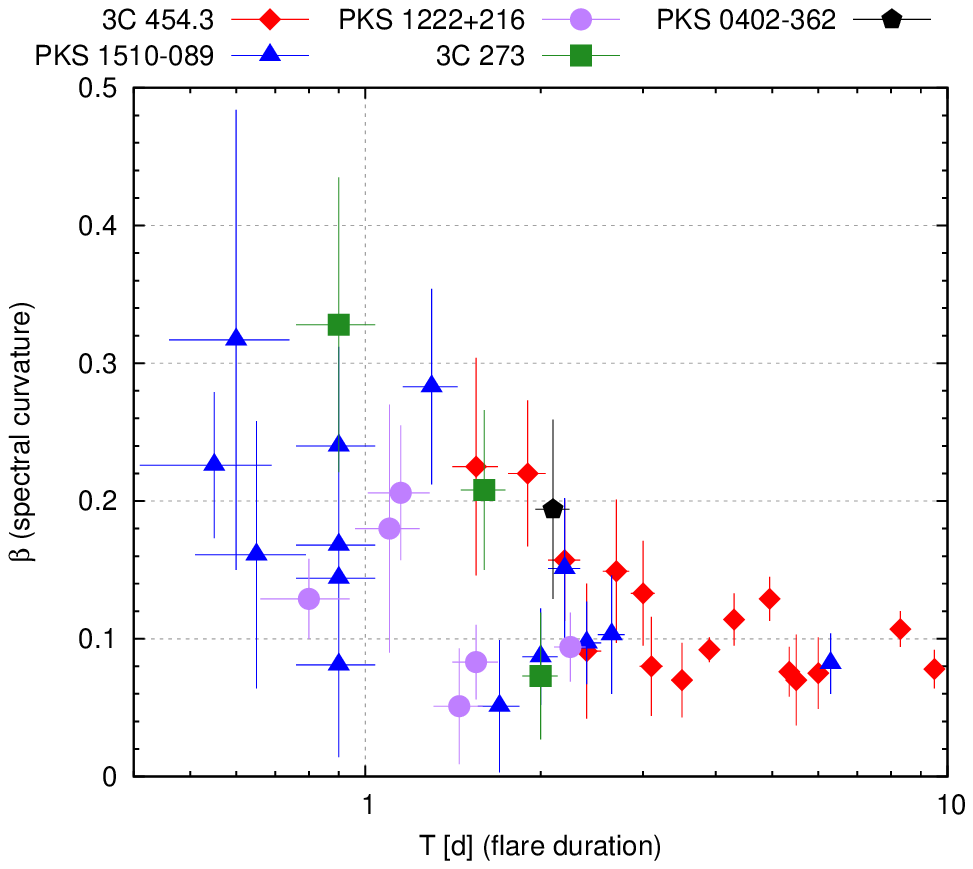}
\includegraphics[width=0.9\columnwidth]{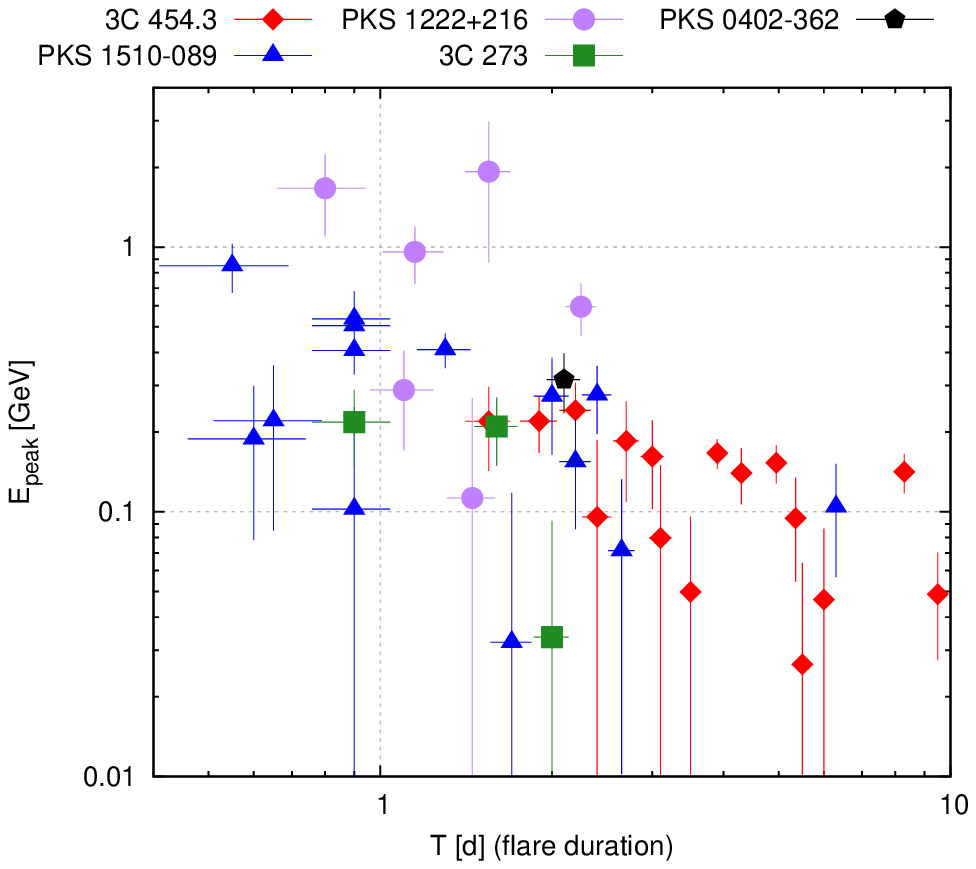}
\includegraphics[width=0.9\columnwidth]{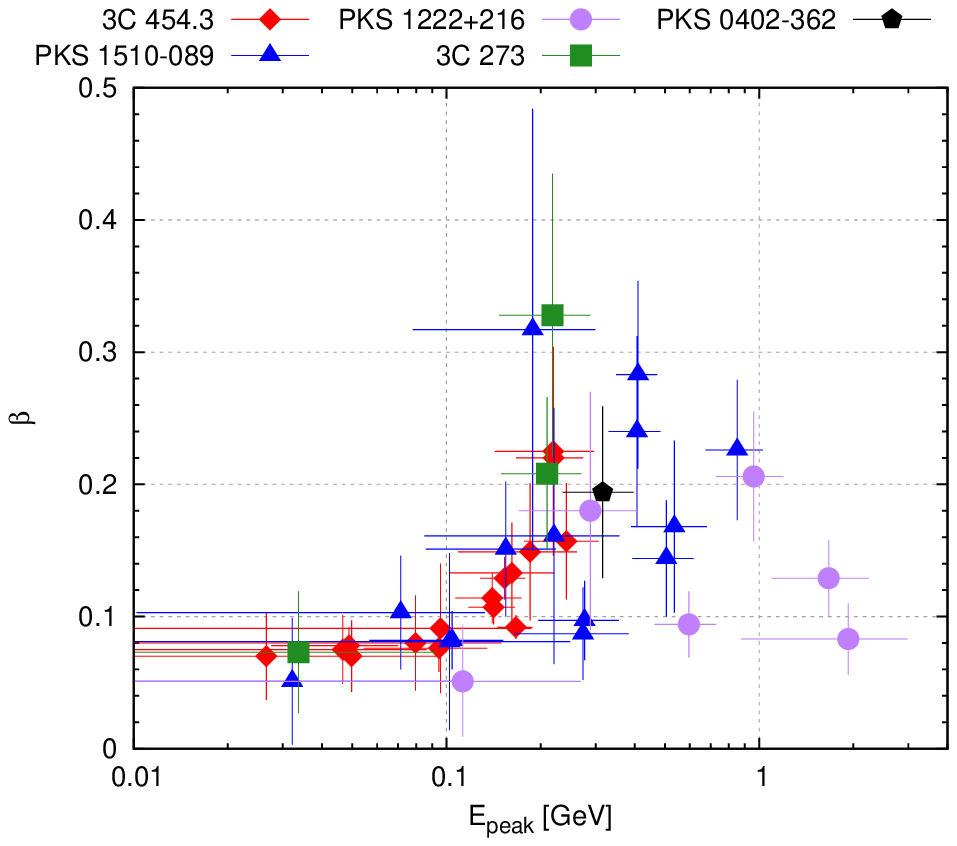}
\caption{Distributions of the spectral curvature $\beta$ and the spectral peak position $E_{\rm peak}$ (resulting from a log parabola fit to the spectrum) versus the duration $T$ (days) of the flare, plotted for the 40 brightest \emph{Fermi}/LAT gamma-ray flares. The shape and color of each point indicate the host blazar.}
\label{fig:beta}
\end{figure}

In Figure \ref{fig:beta}, we show the distributions of $\beta$ and $E_{\rm peak}$ vs. the flare duration $T$. We find that flare duration has a strong influence on the shape of the gamma-ray spectrum. Flares longer than $\simeq 2.5\;{\rm d}$ have gently curved spectra with $\beta \sim 0.1$ and $E_{\rm peak} \lesssim 0.1\;{\rm GeV}$, whereas shorter flares can have a stronger curvature with $E_{\rm peak} > 0.1\;{\rm GeV}$. As was noted in \citetalias{Nalewajko2013}, most of the long flares were produced by 3C~454.3, and they also tend to have more consistent average photon indices $\left<\Gamma\right> \simeq 2.3$, and more symmetric distribution of the time asymmetry parameter. On the other hand, the short flares are more typical for blazars PKS~1510-089 and PKS~1222+216. The latter source stands out by producing flares with the highest $E_{\rm peak}$ values. For completeness, in Figure \ref{fig:beta} we also show the distribution of $\beta$ vs. $E_{\rm peak}$. The spectra with $E_{\rm peak} < 0.1\;{\rm GeV}$ by necessity are characterized by low values of $\beta$. The highest values of $\beta$ are obtained when $E_{\rm peak} \sim 0.2\;{\rm GeV}$.

\section{Discussion}
\label{sec_dis}

In this work we focus on the gamma-ray spectra of blazars integrated during the highest observed gamma-ray fluxes on relatively short time scales of $T < 10\;{\rm d}$. This is a consequence of the flare definition adopted in \citetalias{Nalewajko2013}, and this approach distinguishes this work from most studies of gamma-ray spectra of blazars that focus on maximizing the photon statistics by integrating the spectra on much longer time scales (months -- years). The results of studies performed on longer time scales may not be applicable on shorter time scales. A glance at Figure \ref{fig:top40} reveals many irregularities that are absent in the neat long-term results presented, e.g., by \cite{Abdo2010a}. In the face of such irregularities, we should not expect that these spectra can be well fit by simple spectral models like SPL, BPL or LP. Instead, we should expect at most to produce better or worse approximations of the real spectra with these models.

Of course, one should carefully consider whether these irregularities may be due to any statistical or systematical errors in the analysis of the \emph{Fermi}/LAT data. From the standard maximum likelihood analysis, we know how many reconstructed photons contribute to each flux measurement, and so we are not concerned about the photon statistics. The systematic errors are less understood, and the \emph{Fermi} Collaboration provides only crude estimates based on the in-orbit calibration studies \citep{Ackermann2012}, but these values ($\sim 10\%$) are much lower than the observed amplitudes of the spectral fluctuations. That the power density spectra of bright blazars are power laws without any breaks \citep{Abdo2010c}, including 3C~454.3 in the high state \citep{Ackermann2010}, suggests that flux measurements are equally accurate at all relevant time scales. We will therefore assume that the observed spectral fluctuations are a real property of blazars, and not instrumental artifacts.

Our systematic and unbiased search for the occurrence of spectral breaks returned results that can be characterized as random. The broad distributions of break energies $E_{\rm br}$ and break amounts $\Delta\Gamma$, and the lack of clear correlations with other flare parameters, suggests that there is no unique physical mechanism behind them. In the double absorber model \citep{Poutanen2010}, spectral breaks should be observed at consistent break energy $E_{\rm br} \simeq 5\;{\rm GeV}$ in the source frame at all times, unless the gamma-ray radiation is produced far outside the broad-line region. Our results do not indicate any preference for this $E_{\rm br}$ value, which is consistent with the results of \cite{Harris2012}. While the uncertainty of the intrinsic gamma-ray spectrum of blazars may affect the expected position of spectral breaks produced by absorption, it is rather unlikely that this effect could explain the very broad distribution of $E_{\rm br}$ values shown in Figure \ref{fig:ebr}. The irregularity of the break parameters suggests that they reflect the random spectral fluctuations observed in the binned spectra. At much longer integration time scales, more regular spectral breaks could arise due to non-uniform statistics of such fluctuations. Also, longer integration times are required in order to address the question of $\sim 20\;{\rm GeV}$ breaks arising from absorption by the hydrogen Ly-$\alpha$ continuum.

The main finding of this work is that the spectra of long flares ($T > 2\;{\rm d}$) are more regular than the spectra of short flares ($T < 2\;{\rm d}$), which is illustrated by the distribution of the parameters $\beta$ and $E_{\rm peak}$ of the log-parabola fits to the individual spectra (Fig. \ref{fig:beta}). The short flares often have their spectral peak within the \emph{Fermi}/LAT band ($E_{\rm peak} > 0.1\;{\rm GeV}$), which is not the case for the long-term average spectra of FSRQ blazars \citep{Abdo2010a}. This may have profound implications for the theoretical picture of dissipation and particle acceleration in relativistic AGN jets. The regular (gently broken power-law) gamma-ray spectra of blazars observed on long time scales may generally be superpositions of many simple components. Each of those components may have a narrowly peaked particle energy distribution and produce a short subflare contributing to the overall observed light curve. If the emission of such a component is coherent in photon energy and time, it may also be coherent in space. In short, we propose that the emitting regions of blazars are macroscopically turbulent, with individual eddies producing short radiation spikes of narrow spectrum.

Of course, it is not a new idea that blazar jets are fundamentally turbulent \citep{Begelman1984,Jones1988}. Observations that the power density spectra of blazar light curves are consistent with simple power laws \citep{Abdo2010c} provide strong evidence of this. Dedicated studies have been done in order to explain the observed blazar light curves in terms of emission produced in relativistic turbulence \citep{Marscher2014,Calafut2014}. What we demonstrate here is that \emph{Fermi}/LAT is able to resolve these turbulent fluctuations in energy when observations are integrated on time scales $T < 2\;{\rm d}$. If the maximum size of the turbulent eddies is limited by the jet radius $R$ at distance $r$, the observed time scale of a single fluctuation is limited by $t_{\rm var} \lesssim R/(\mathcal{D}c) \simeq r/(\Gamma^2c) \simeq 3\;{\rm d}\;(r/1\;{\rm pc})(\Gamma/20)^{-2}$, where we assumed that $R = \theta r$, $\theta \simeq 1/\Gamma$ is the jet opening angle, $\mathcal{D} \simeq \Gamma$ is the Doppler factor, and $\Gamma$ is the jet Lorentz factor. If the gamma-ray radiation in blazars is produced mainly at distance scale $r \sim 0.1 - 1\;{\rm pc}$ \citep{Nalewajko2014}, this time scale would be broadly consistent with our results.

Macroscopic turbulence may be generated in relativistic strongly-magnetized jets by current-driven instability (CDI). Global 3D general relativistic simulations of black-hole jets suggest that they can be marginally stable to global CDI modes \citep{McKinney2009}. For sufficient dissipation efficiency, the most desirable situation is to have a jet stable globally and unstable locally \citep{Begelman1998}. Recent numerical results demonstrate that this is theoretically feasible \citep{Porth2014}.
Other mechanisms that can produce macroscopic turbulence in relativistic jets include plasma heating in the jet boundary layer resulting from interaction with the environment \citep{Kohler2015}, or magnetic reconnection generating relativistic bulk motions and/or relativistic heating \citep{Giannios09}.

If the broad spectra of blazars are a superposition of many narrow components that can be resolved on daily time scales, there is no need to explain the regular long-term average spectra by microscopic processes like diffusive shock acceleration at relativistic shock waves \citep{Spitkovsky2008}. Instead, one can consider a process that energizes many particles to a similar energy, like relativistic magnetic reconnection. Recent numerical studies of relativistic reconnection \citep{Sironi2014,Guo2014,Werner14} demonstrate that for sufficiently high magnetization $\sigma \gtrsim 10$, the particles form a very hard power-law distribution $N(\gamma) \propto \gamma^{-p}$ with $p < 2$, therefore, most energy is contained (and can be radiated) at the high-energy end of the distribution defined by the total energy content in the system.

The gamma-ray spectra of flaring blazars presented here can be used to constrain the underlying energy distribution of ultra-relativistic electrons, under the assumption that the gamma-ray radiation is produced by inverse Compton scattering of external photon fields \citep{Sikora2009}. However, the relation between the energy distribution of electrons and the spectrum of radiation scattered by them is complicated; for example, even in the Thomson regime the curvature parameter for the electrons is about four times greater than the curvature parameter for the radiation \citep{Massaro2006}. Knowledge of the simultaneous synchrotron radiation spectra in the NIR/optical/UV band and detailed modeling of the broad-band blazar spectra would be necessary in order to determine the electron energy distribution. Our proposition for the turbulent origin of the gamma-ray spectra of blazars further complicates this problem by requiring a multi-zone approach.

In \citetalias{Nalewajko2013}, a dichotomy was revealed between the temporal properties of the brightest gamma-ray flares of blazars. On one hand, most flares produced by 3C~454.3 are long, with complex light curves (multiple subflares of comparable peak flux), without clear time asymmetry. On the other hand, most flares produced by PKS 1510-089 and PKS 1222+216 are short, with simple light curves, and a tendency for the flux decay time scale to be longer than the flux raising time scale. This dichotomy was suggested to be an observational effect, with the viewing angle of the jet much smaller in the case of 3C~454.3, resulting in the more uniform Doppler beaming of all emitting regions. Now we add to this picture the systematic differences between the observed gamma-ray spectra. The spectra of long flares are more regular than the spectra of short flares, as the former consist of more elementary narrow components. The interpretation of the source dichotomy in terms of the viewing angle is consistent with this, as more uniform Doppler beaming (smaller viewing angle) is required to observe more spectral components at comparable flux levels (hence more regular integrated spectra).

Two of the sources contributing to the sample of flares studied here, PKS 1222+216 and PKS 1510-089, were detected in the VHE gamma-ray band (above 70~GeV) by ground-based Cherenkov telescopes. The \emph{MAGIC} observation of PKS~1222+216 at MJD~55364.9 \citep{Aleksic2011} coincided with the beginning of Flare \#12 \citep{Tanaka2011}. The \emph{H.E.S.S.} observations of PKS~1510-089 at MJD~54910-54918 \citep{HESS2013} coincided with Flare \#20, and \emph{MAGIC} observations of PKS~1510-089 between MJD~55975 and MJD~56020 \citep{Aleksic2014} coincided with a series of flares: Flares \#19, \#29, \#28, and \#32. The observed $\nu F_\nu$ flux of PKS~1222+216 at 100~GeV was $\sim~10^{-10}\;{\rm erg\,s^{-1}\,cm^{-2}}$, while that of PKS~1510-089 was much lower, $\sim~10^{-11}\;{\rm erg\,s^{-1}\,cm^{-2}}$. The \emph{Fermi}/LAT spectra of the flares coinciding with the VHE detections do not appear to be different from the spectra of other flares for the same sources. Unfortunately, because of the very limited photon statistics above 10~GeV at short integration times, we cannot determine which of these spectra extend to the VHE band at detectable levels. However, by extrapolating the best-fit log-parabola models (and ignoring the problem of possible absorption of the VHE gamma-rays by intrinsic and external soft radiation fields), we can make an educated guess which of the flares were most likely to be detected at 100~GeV at the high level of $\nu F_\nu \gtrsim 10^{-10}\;{\rm erg\,s^{-1}\,cm^{-2}}$: for PKS~1222+216, Flares \#8, \#12, \#26 and \#35; for PKS~1510-089, Flare \#11; and for 3C~454.3, Flare \#1. PKS~1222+216 appears to be most likely to be detected in the VHE band due to its consistently hard gamma-ray spectrum \citep{Nalewajko2013}, although so far --- to our knowledge --- it was detected only once. In the case of 3C~454.3, at redshift $z = 0.859$, any VHE detection would be extraordinarily significant in constraining the extragalactic background light (EBL), and we suggest that it should be vigorously pursued by the current (\emph{H.E.S.S.}, \emph{MAGIC} and \emph{VERITAS}) and future (\emph{CTA}) VHE observatories.

\section{Conclusions}
\label{sec_con}

We performed a spectral analysis with the \emph{Fermi}/LAT of the sample (selected in \citetalias{Nalewajko2013}) of the 40 brightest gamma-ray flares of blazars (FSRQs) detected in the first four years of the \emph{Fermi} mission. The gamma-ray spectra are integrated over relatively short time scales $T < 10\;{\rm d}$, and they show significant and variable departures from the long-term average spectra of the same sources. We performed a uniform search for the occurrence of spectral breaks. The break energies show a broad distribution and no preference for the fixed value of 5 GeV in the source frame predicted by the double-absorber model of \cite{Poutanen2010}. In order to compare the basic structures of the observed spectra, we fitted them with a log parabola model and found an interesting trend of the model parameters with the flare duration. Short flares ($T < 2\;{\rm d}$) often show a strong spectral curvature with the SED peak within the \emph{Fermi}/LAT range $E_{\rm peak} > 0.1\;{\rm GeV}$, while all long flares show a mild spectral curvature with a SED peak below the LAT range. The dichotomy between typical properties of flares observed in sources 3C~454.3 vs. PKS~1510-089 and others, first described in \citetalias{Nalewajko2013}, is extended to include differences between the observed gamma-ray spectra. The extrapolation of the log-parabola spectral models to $100\;{\rm GeV}$ identifies several candidates for VHE detections.

We suggest that the irregular gamma-ray spectra observed by the \emph{Fermi}/LAT for short blazar flares reflect macroscopic turbulence in relativistic jets with individual emitting regions (turbulent eddies) having a narrow energy distribution of emitting particles. The superposition of many such spectral components peaking at different energies would then result in the regular power-law spectra observed over long time scales. This would eliminate the need for localized stochastic particle acceleration in the blazar zones of relativistic AGN jets. Further investigation is necessary in order to understand the underlying energy distribution of ultra-relativistic electrons.

\section*{Acknowledgements}

The authors thank the anonymous referee, as well as Greg Madejski and Aneta Siemiginowska, Mitch Begelman and Phil Armitage for helpful comments.
This work is based on the publicly available data from the \emph{Fermi} Large Area Telescope operated by NASA and Department of Energy in collaboration with institutions from France, Italy, Japan, and Sweden.
S.K. was supported by NSF grant AST-0907872, NASA Astrophysics Theory Program grant NNX09AG02G, and NASA's \emph{Fermi} Gamma-ray Space Telescope Guest Investigator program.
K.N. was supported by NASA through Einstein Postdoctoral Fellowship grant number PF3-140130 awarded by the Chandra X-ray Center, which is operated by the Smithsonian Astrophysical Observatory for NASA under contract NAS8-03060.

\clearpage

\clearpage

\clearpage

\begin{table*}
\centering
\renewcommand{\arraystretch}{1.5}
\caption{Statistically significant breaks discovered in the spectra of the 40 brightest \emph{Fermi}/LAT gamma-ray flares. Breaks were recorded if the spectrum was better-fit by a BPL rather than SPL or LP model, with a $\Delta$AIC $>2$ for the BPL compared to the next-best fit. ${\rm MJD}_{\rm peak}$ is the moment of flux peak, $F_{\rm peak}$ is the peak photon flux in units of $10^{-6}\;{\rm ph\,s^{-1}\,cm^{-2}}$, $T$ is the duration of the flare in days, $E_{\rm br}$ (obs) and $E_{\rm br}$ (source) are the break energies, in GeV, in the observer and source frame respectively, and $\Gamma_1$ and $\Gamma_2$ are the spectral indices on either side of the break energy. The break classification identifies the break as resulting from a fit in the entire energy range $0.1 - 10$ GeV (``primary'') or resulting from a fit only in the energy range below or above the primary break (``secondary: low'' and ``secondary: high'').
}
\label{tab:sigbreaks}
\begin{tabular}{llllrrrrrr}
\hline
\# & Blazar & $\rm MJD_{peak}$ & Break classification & $F_{peak}$ & $T$ & $E_{\rm br}$ (obs) & $E_{\rm br}$ (source) & $\Gamma_1$ & $\Gamma_2$ \\
\hline
1 & 3C 454.3 & 55520.0 & secondary: low & 77.2 $\pm$ 2.4 & 3.9 & $0.35_{-0.06}^{+0.05}$ & $0.65_{-0.11}^{+0.09}$ & 2.03 $\pm$ 0.03 & 2.25 $\pm$ 0.04 \\
1 & 3C 454.3 & 55520.0 & secondary: high & 77.2 $\pm$ 2.4 & 3.9 & $6.46_{-2.07}^{+0.62}$ & $12.00_{-3.85}^{+1.15}$ & 2.43 $\pm$ 0.10 & 3.59 $\pm$ 0.31 \\
4 & 3C 454.3 & 55550.3 & secondary: high & 23.6 $\pm$ 1.3 & 8.3 & $8.12_{-2.47}^{+3.79}$ & $15.09_{-4.59}^{+7.05}$ & 2.54 $\pm$ 0.05 & 3.32 $\pm$ 0.43 \\
5 & 3C 454.3 & 55167.8 & secondary: low & 21.8 $\pm$ 1.5 & 4.3 & $0.15_{-0.02}^{+0.01}$ & $0.28_{-0.04}^{+0.02}$ & 1.62 $\pm$ 0.27 & 2.22 $\pm$ 0.05 \\
6 & 3C 454.3 & 55567.8 & secondary: high & 19.2 $\pm$ 1.6 & 5.4 & $5.82_{-2.22}^{+1.81}$ & $10.81_{-4.13}^{+3.37}$ & 2.36 $\pm$ 0.05 & 3.06 $\pm$ 0.38 \\
7 & 3C 454.3 & 55301.5 & primary & 16.4 $\pm$ 1.8 & 3.5 & $1.64_{-0.32}^{+0.51}$ & $3.05_{-0.60}^{+0.96}$ & 2.25 $\pm$ 0.04 & 2.89 $\pm$ 0.20 \\
8 & PKS 1222+216 & 55316.6 & secondary: high & 15.6 $\pm$ 0.9 & 0.8 & $7.16_{-0.80}^{+1.58}$ & $10.26_{-1.14}^{+2.26}$ & 1.97 $\pm$ 0.17 & 3.46 $\pm$ 0.64 \\
10 & 3C 454.3 & 55294.1 & secondary: high & 15.3 $\pm$ 0.8 & 9.5 & $8.35_{-2.68}^{+1.12}$ & $15.52_{-4.98}^{+2.08}$ & 2.59 $\pm$ 0.09 & 4.13 $\pm$ 0.81 \\
12 & PKS 1222+216 & 55365.8 & primary & 14.2 $\pm$ 1.0 & 2.3 & $0.56_{-0.05}^{+0.75}$ & $0.80_{-0.07}^{+1.08}$ & 1.76 $\pm$ 0.08 & 2.27 $\pm$ 0.06 \\
12 & PKS 1222+216 & 55365.8 & secondary: low & 14.2 $\pm$ 1.0 & 2.3 & $0.17_{-0.02}^{+0.01}$ & $0.24_{-0.03}^{+0.01}$ & 2.76 $\pm$ 0.36 & 1.58 $\pm$ 0.13 \\
13 & 3C 454.3 & 55163.1 & primary & 11.8 $\pm$ 1.4 & 3.0 & $1.37_{-0.28}^{+0.27}$ & $2.55_{-0.52}^{+0.51}$ & 2.17 $\pm$ 0.06 & 3.19 $\pm$ 0.23 \\
14 & 3C 454.3 & 55305.5 & primary & 11.1 $\pm$ 1.5 & 1.9 & $0.56_{-0.06}^{+0.22}$ & $1.03_{-0.10}^{+0.41}$ & 1.98 $\pm$ 0.09 & 2.94 $\pm$ 0.15 \\
20 & PKS 1510--089 & 54917.0 & primary & 9.9 $\pm$ 0.7 & 2.4 & $1.50_{-0.28}^{+0.30}$ & $2.04_{-0.38}^{+0.41}$ & 2.03 $\pm$ 0.05 & 2.75 $\pm$ 0.17 \\
21 & PKS 1510--089 & 55851.9 & primary & 9.9 $\pm$ 1.0 & 0.7 & $1.05_{-0.29}^{+0.25}$ & $1.42_{-0.39}^{+0.33}$ & 2.04 $\pm$ 0.15 & 3.24 $\pm$ 0.48 \\
22 & 3C 454.3 & 55195.2 & primary & 9.7 $\pm$ 0.8 & 2.2 & $1.97_{-0.30}^{+0.49}$ & $3.66_{-0.56}^{+0.92}$ & 2.12 $\pm$ 0.06 & 3.68 $\pm$ 0.46 \\
25 & 3C 454.3 & 55327.2 & primary & 8.8 $\pm$ 1.2 & 3.1 & $1.80_{-0.33}^{+0.35}$ & $3.34_{-0.62}^{+0.66}$ & 2.23 $\pm$ 0.06 & 2.98 $\pm$ 0.26 \\
26 & PKS 1222+216 & 55342.1 & secondary: high & 8.7 $\pm$ 0.8 & 1.6 & $1.93_{-0.71}^{+1.46}$ & $2.77_{-1.02}^{+2.09}$ & 1.82 $\pm$ 0.08 & 2.24 $\pm$ 0.14 \\
30 & PKS 1510--089 & 54961.8 & primary & 8.2 $\pm$ 0.7 & 0.6 & $0.56_{-0.11}^{+0.22}$ & $0.76_{-0.15}^{+0.30}$ & 2.04 $\pm$ 0.24 & 3.56 $\pm$ 0.49 \\
33 & 3C 454.3 & 55154.8 & primary & 7.8 $\pm$ 0.9 & 2.4 & $0.27_{-0.04}^{+0.04}$ & $0.50_{-0.07}^{+0.08}$ & 1.79 $\pm$ 0.20 & 2.50 $\pm$ 0.09 \\
35 & PKS 1222+216 & 55377.5 & primary & 7.6 $\pm$ 0.7 & 1.5 & $0.35_{-0.23}^{+0.12}$ & $0.51_{-0.33}^{+0.17}$ & 1.86 $\pm$ 0.17 & 2.33 $\pm$ 0.10 \\
37 & PKS 1510--089 & 55876.1 & secondary: high & 7.4 $\pm$ 0.8 & 1.7 & $0.61_{-0.10}^{+0.53}$ & $0.83_{-0.14}^{+0.72}$ & 4.54 $\pm$ 0.90 & 2.31 $\pm$ 0.16 \\
38 & PKS 1222+216 & 55234.0 & primary & 7.4 $\pm$ 0.8 & 1.1 & $0.21_{-0.02}^{+0.03}$ & $0.29_{-0.02}^{+0.05}$ & 0.50 $\pm$ 0.59 & 2.49 $\pm$ 0.13 \\
40 & 3C 454.3 & 55091.6 & primary & 7.1 $\pm$ 0.8 & 5.5 & $3.09_{-0.84}^{+0.76}$ & $5.75_{-1.57}^{+1.41}$ & 2.35 $\pm$ 0.05 & 4.63 $\pm$ 0.97
\\
\hline
\end{tabular}
\end{table*}

\begin{table*}
\centering
\caption{Log-parabola fit parameters for the brightest gamma-ray flares of blazars. ${\rm MJD_{peak}}$ is the moment of flux peak, $F_{\rm peak}$ is the peak photon flux in units of $10^{-6}\;{\rm ph\,s^{-1}\,cm^{-2}}$, $t_1$ is the flux doubling time scale, $t_2$ is the flux halving time scale, $T=(t_1+t_2)$ is the flare duration, $\alpha$ is the spectral index at the pivot energy, $\beta$ is the spectral curvature, and $E_{\rm peak}$ is the implied energy of the observer-frame $\nu F_\nu$ spectral peak in units of GeV. All times are in units of days.
}
\label{tab:logpar}
\begin{tabular}{rllrrrrrrrr}
\hline
\# & Blazar        & ${\rm MJD_{peak}}$ & ${F_{\rm peak}}$     & $t_{1}$ & $t_2$ & $T$ & $\alpha$        & $\beta$         & $E_{\rm peak}$      \\
\hline
1  & 3C 454.3     & 55520.0     & 77.2 $\pm$ 2.4 & 2.9     & 1.0   & 3.9 & 2.20 $\pm$ 0.01 & 0.09 $\pm$ 0.01 & 0.17 $\pm$ 0.02 \\
2  & 3C 454.3     & 55526.9     & 33.3 $\pm$ 1.8 & 1.1     & 3.9   & 5.0 & 2.31 $\pm$ 0.02 & 0.13 $\pm$ 0.02 & 0.15 $\pm$ 0.03 \\
3  & PKS 1510--089 & 55853.8     & 26.6 $\pm$ 2.0 & 0.2     & 0.4   & 0.6 & 1.76 $\pm$ 0.08 & 0.23 $\pm$ 0.05 & 0.85 $\pm$ 0.18 \\
4  & 3C 454.3     & 55550.3     & 23.6 $\pm$ 1.3 & 3.2     & 5.1   & 8.3 & 2.27 $\pm$ 0.02 & 0.11 $\pm$ 0.01 & 0.14 $\pm$ 0.02 \\
5  & 3C 454.3     & 55167.8     & 21.8 $\pm$ 1.5 & 1.2     & 3.1   & 4.3 & 2.29 $\pm$ 0.03 & 0.11 $\pm$ 0.02 & 0.14 $\pm$ 0.04 \\
6  & 3C 454.3     & 55567.8     & 19.2 $\pm$ 1.6 & 2.9     & 2.5   & 5.4 & 2.25 $\pm$ 0.02 & 0.08 $\pm$ 0.02 & 0.09 $\pm$ 0.04 \\
7  & 3C 454.3     & 55301.5     & 16.4 $\pm$ 1.8 & 1.3     & 2.2   & 3.5 & 2.32 $\pm$ 0.04 & 0.07 $\pm$ 0.03 & \textless 0.10  \\
8  & PKS 1222+216 & 55316.6     & 15.6 $\pm$ 0.9 & 0.4     & 0.4   & 0.8 & 1.69 $\pm$ 0.05 & 0.13 $\pm$ 0.03 & 1.67 $\pm$ 0.56 \\
9  & PKS 1510--089 & 55872.8     & 15.3 $\pm$ 1.2 & 0.2     & 0.7   & 0.9 & 2.00 $\pm$ 0.06 & 0.14 $\pm$ 0.04 & 0.51 $\pm$ 0.11 \\
10 & 3C 454.3     & 55294.1     & 15.3 $\pm$ 0.8 & 5.9     & 3.6   & 9.5 & 2.36 $\pm$ 0.02 & 0.08 $\pm$ 0.01 & 0.05 $\pm$ 0.02 \\
11 & PKS 1510--089 & 55867.8     & 14.2 $\pm$ 1.5 & 0.5     & 1.5   & 2.0 & 2.11 $\pm$ 0.06 & 0.09 $\pm$ 0.04 & 0.27 $\pm$ 0.12 \\
12 & PKS 1222+216 & 55365.8     & 14.2 $\pm$ 1.0 & 1.0     & 1.3   & 2.3 & 1.97 $\pm$ 0.04 & 0.09 $\pm$ 0.03 & 0.60 $\pm$ 0.14 \\
13 & 3C 454.3     & 55163.1     & 11.8 $\pm$ 1.4 & 2.0     & 1.0   & 3.0 & 2.30 $\pm$ 0.05 & 0.13 $\pm$ 0.04 & 0.16 $\pm$ 0.07 \\
14 & 3C 454.3     & 55305.5     & 11.1 $\pm$ 1.5 & 0.8     & 1.1   & 1.9 & 2.36 $\pm$ 0.06 & 0.22 $\pm$ 0.05 & 0.22 $\pm$ 0.05 \\
15 & PKS 1510--089 & 55746.2     & 11.1 $\pm$ 1.4 & 0.4     & 0.5   & 0.9 & 2.26 $\pm$ 0.09 & 0.08 $\pm$ 0.07 & \textless 0.25  \\
16 & PKS 1510--089 & 54948.0     & 10.6 $\pm$ 2.3 & 0.9     & 1.3   & 2.2 & 2.35 $\pm$ 0.06 & 0.15 $\pm$ 0.05 & 0.16 $\pm$ 0.07 \\
17 & 3C 273       & 55095.3     & 10.3 $\pm$ 0.9 & 0.4     & 1.6   & 2.0 & 2.39 $\pm$ 0.06 & 0.07 $\pm$ 0.05 & \textless 0.09  \\
18 & 3C 273       & 55090.5     & 10.2 $\pm$ 0.9 & 0.4     & 1.2   & 1.6 & 2.36 $\pm$ 0.07 & 0.21 $\pm$ 0.06 & 0.21 $\pm$ 0.06 \\
19 & PKS 1510--089 & 55980.6     & 10.0 $\pm$ 1.3 & 0.5     & 0.8   & 1.3 & 2.11 $\pm$ 0.08 & 0.28 $\pm$ 0.07 & 0.41 $\pm$ 0.06 \\
20 & PKS 1510--089 & 54917.0     & 9.9 $\pm$ 0.7  & 1.9     & 0.5   & 2.4 & 2.12 $\pm$ 0.04 & 0.10 $\pm$ 0.03 & 0.28 $\pm$ 0.08 \\
21 & PKS 1510--089 & 55851.9     & 9.9 $\pm$ 1.0  & 0.2     & 0.5   & 0.7 & 2.26 $\pm$ 0.12 & 0.16 $\pm$ 0.10 & 0.22 $\pm$ 0.14 \\
22 & 3C 454.3     & 55195.2     & 9.7 $\pm$ 0.8  & 0.6     & 1.6   & 2.2 & 2.23 $\pm$ 0.06 & 0.16 $\pm$ 0.04 & 0.24 $\pm$ 0.06 \\
23 & PKS 1222+216 & 55310.7     & 9.6 $\pm$ 0.8  & 0.5     & 0.7   & 1.2 & 1.73 $\pm$ 0.08 & 0.21 $\pm$ 0.05 & 0.96 $\pm$ 0.24 \\
24 & 3C 454.3     & 55323.5     & 9.1 $\pm$ 0.9  & 4.2     & 1.8   & 6.0 & 2.36 $\pm$ 0.04 & 0.08 $\pm$ 0.03 & 0.05 $\pm$ 0.04 \\
25 & 3C 454.3     & 55327.2     & 8.8 $\pm$ 1.2  & 1.9     & 1.2   & 3.1 & 2.29 $\pm$ 0.05 & 0.08 $\pm$ 0.04 & 0.08 $\pm$ 0.08 \\
26 & PKS 1222+216 & 55342.1     & 8.7 $\pm$ 0.8  & 0.8     & 0.8   & 1.6 & 1.78 $\pm$ 0.05 & 0.08 $\pm$ 0.03 & 1.93 $\pm$ 1.16 \\
27 & 3C 273       & 55202.9     & 8.7 $\pm$ 1.1  & 0.3     & 0.6   & 0.9 & 2.54 $\pm$ 0.12 & 0.33 $\pm$ 0.11 & 0.22 $\pm$ 0.07 \\
28 & PKS 1510--089 & 55990.8     & 8.5 $\pm$ 0.7  & 5.2     & 1.1   & 6.3 & 2.26 $\pm$ 0.03 & 0.08 $\pm$ 0.02 & 0.10 $\pm$ 0.05 \\
29 & PKS 1510--089 & 55983.1     & 8.4 $\pm$ 0.8  & 0.5     & 0.4   & 0.9 & 2.10 $\pm$ 0.09 & 0.24 $\pm$ 0.07 & 0.41 $\pm$ 0.08 \\
30 & PKS 1510--089 & 54961.8     & 8.2 $\pm$ 0.7  & 0.3     & 0.3   & 0.6 & 2.62 $\pm$ 0.18 & 0.32 $\pm$ 0.17 & 0.19 $\pm$ 0.11 \\
31 & 3C 454.3     & 55214.3     & 8.0 $\pm$ 0.9  & 0.8     & 0.8   & 1.6 & 2.37 $\pm$ 0.09 & 0.23 $\pm$ 0.08 & 0.22 $\pm$ 0.08 \\
32 & PKS 1510--089 & 56002.4     & 7.8 $\pm$ 1.0  & 1.8     & 0.9   & 2.7 & 2.40 $\pm$ 0.06 & 0.10 $\pm$ 0.04 & 0.07 $\pm$ 0.06 \\
33 & 3C 454.3     & 55154.8     & 7.8 $\pm$ 0.9  & 0.5     & 1.9   & 2.4 & 2.30 $\pm$ 0.06 & 0.09 $\pm$ 0.05 & 0.10 $\pm$ 0.10 \\
34 & PKS 1510--089 & 55767.6     & 7.6 $\pm$ 0.8  & 0.4     & 0.5   & 0.9 & 1.98 $\pm$ 0.09 & 0.17 $\pm$ 0.07 & 0.54 $\pm$ 0.14 \\
35 & PKS 1222+216 & 55377.5     & 7.6 $\pm$ 0.7  & 0.4     & 1.1   & 1.5 & 2.15 $\pm$ 0.07 & 0.05 $\pm$ 0.04 & \textless 0.26  \\
36 & 3C 454.3     & 55283.9     & 7.6 $\pm$ 1.1  & 1.3     & 1.4   & 2.7 & 2.30 $\pm$ 0.07 & 0.15 $\pm$ 0.05 & 0.19 $\pm$ 0.08 \\
37 & PKS 1510--089 & 55876.1     & 7.4 $\pm$ 0.8  & 0.7     & 1.0   & 1.7 & 2.28 $\pm$ 0.07 & 0.05 $\pm$ 0.05 & \textless 0.12  \\
38 & PKS 1222+216 & 55234.0     & 7.4 $\pm$ 0.8  & 0.7     & 0.4   & 1.1 & 2.20 $\pm$ 0.11 & 0.18 $\pm$ 0.09 & 0.29 $\pm$ 0.12 \\
39 & PKS 0402--362 & 55827.5     & 7.3 $\pm$ 1.0  & 0.3     & 1.8   & 2.1 & 2.18 $\pm$ 0.08 & 0.19 $\pm$ 0.07 & 0.32 $\pm$ 0.09 \\
40 & 3C 454.3     & 55091.6     & 7.1 $\pm$ 0.8  & 4.4     & 1.1   & 5.5 & 2.41 $\pm$ 0.04 & 0.07 $\pm$ 0.03 & \textless 0.07 \\
\hline
\end{tabular}
\label{lastpage}
\end{table*}

\clearpage


\begin{thebibliography}{0}

\bibitem[{Abdo} {et~al.}(2009)]{Abdo2009}
{Abdo}, A.~A., {et~al.}, 2009, ApJ, 699, 817

\bibitem[{Abdo} {et~al.}(2010a)]{Abdo2010a}
{Abdo}, A.~A., {Ackermann}, M., {Ajello}, M., {et~al.} 2010a, ApJ, 710, 1271

\bibitem[Abdo et~al.(2010b)]{Abdo2010b}
Abdo, A.~A., Ackermann, M., Agudo, I., et~al., 2010b, ApJ, 716, 30

\bibitem[Abdo et~al.(2010c)]{Abdo2010c}
Abdo, A.~A., Ackermann, M., Ajello, M., et~al. 2010c, ApJ, 722, 520

\bibitem[{Abdo} {et~al.}(2011)]{Abdo2011}
{Abdo}, A.~A., {Ackermann}, M., {Ajello}, M., {et~al.}, 2011, ApJ, 733, L26

\bibitem[{Ackermann} {et~al.}(2010)]{Ackermann2010}
{Ackermann}, M., {Ajello}, M., {Baldini}, L., {et~al.} 2010, ApJ, 721, 1383

\bibitem[Ackermann et~al.(2011)]{Ackermann2011}
Ackermann, M., Ajello, M., Allafort, A., et~al.\ 2011, ApJ, 743, 171

\bibitem[Ackermann et~al.(2012)]{Ackermann2012}
Ackermann, M., Ajello, M., Albert, A., et~al.\ 2012, \apjs, 203, 4

\bibitem[{Aleksi{\'c}} {et~al.}(2011)]{Aleksic2011}
{Aleksi{\'c}}, J., {et~al.}, 2011, ApJ, 730, L8

\bibitem[Aleksi{\'c} et~al.(2014)]{Aleksic2014}
Aleksi{\'c}, J., Ansoldi, S., Antonelli, L.~A., et~al.\ 2014, A\&A, 569, AA46

\bibitem[{Atwood} {et~al.}(2009)]{Atwood2009}
{Atwood}, W.~B., {Abdo}, A.~A., {Ackermann}, M., et~al., 2009, ApJ, 697, 1071

\bibitem[Begelman(1998)]{Begelman1998}
Begelman, M.~C., 1998, ApJ, 493, 291

\bibitem[{Begelman} {et~al.}(1984)]{Begelman1984}
{Begelman}, {M.~C.}, {Blandford}, {R.~D.}, {Rees}, M.~J., 1984, RvMP, 56, 255

\bibitem[{Bozdogan(1987)}]{Bozdogan1987}
Bozdogan, H. 1987, Psychometrika, 52, 345

\bibitem[Calafut \& Wiita(2014)]{Calafut2014}
Calafut, V., \& Wiita, P.~J.\ 2014, arXiv:1407.3687

\bibitem[Cerruti et~al.(2013)]{Cerruti2013}
Cerruti, M., Dermer, C.~D., Lott, B., Boisson, C., \& Zech, A.\ 2013, ApJ, 771, L4

\bibitem[Dermer \& Schlickeiser(1993)]{1993ApJ...416..458D}
Dermer, C.~D., \& Schlickeiser, R.\ 1993, \apj, 416, 458

\bibitem[{Finke} \& {Dermer}(2010)]{Finke2010}
{Finke}, J.~D., \& {Dermer}, C.~D., 2010, ApJ, 714, L303

\bibitem[{Fossati} {et~al.}(1998)]{Fossati1998}
{Fossati}, G., {Maraschi}, L., {Celotti}, A., {Comastri}, A., \& {Ghisellini}, G., 1998, MNRAS, 299, 433

\bibitem[\protect\citeauthoryear{{Giannios}, {Uzdensky} \&
  {Begelman}}{{Giannios} et~al.}{2009}]{Giannios09}
{Giannios} D.,  {Uzdensky} D.~A.,    {Begelman} M.~C.,  2009, \mnras, 395, L29

\bibitem[Guo et~al.(2014)]{Guo2014}
Guo, F., Li, H., Daughton, W., \& Liu, Y.-H.\ 2014, Physical Review Letters, 113, 155005

\bibitem[Harris et~al.(2012)]{Harris2012}
Harris, J., Daniel, M.~K., \& Chadwick, P.~M.\ 2012, ApJ, 761, 2

\bibitem[Harris et al.(2014)]{Harris2014}
Harris, J., Chadwick, P.~M., \& Daniel, M.~K.\ 2014, \mnras, 441, 3591

\bibitem[H.E.S.S.~Collaboration et~al.(2013)]{HESS2013}
H.E.S.S.~Collaboration, Abramowski, A., Acero, F., et al.\ 2013, A\&A, 554, AA107

\bibitem[Jones(1988)]{Jones1988}
Jones, T.~W.\ 1988, ApJ, 332, 678

\bibitem[\protect\citeauthoryear{{Kohler} \& {Begelman}}{{Kohler} \&
  {Begelman}}{2015}]{Kohler2015}
{Kohler} S.,  {Begelman} M.~C.,  2015, \mnras, 446, 1195

\bibitem[{Lewis {et~al.}(2011)Lewis, Butler, \& Gilbert}]{Lewis2011}
Lewis, F., Butler, A., \& Gilbert, L. 2011, Methods in Ecology and Evolution,
  2, 155

\bibitem[Marscher(2014)]{Marscher2014}
Marscher, A.~P.\ 2014, \apj, 780, 87

\bibitem[Massaro et~al.(2006)]{Massaro2006}
Massaro, E., Tramacere, A., Perri, M., Giommi, P., \& Tosti, G.\ 2006, A\&A, 448, 861

\bibitem[{McKinney} \& {Blandford}(2009)]{McKinney2009}
{McKinney}, J.~C., \& {Blandford}, R.~D., 2009, MNRAS, 394, L126

\bibitem[Nalewajko(2013)]{Nalewajko2013}
Nalewajko, K.\ 2013, MNRAS, 430, 1324

\bibitem[Nalewajko et~al.(2014)]{Nalewajko2014}
Nalewajko, K., Begelman, M.~C., \& Sikora, M.\ 2014, ApJ, 789, 161

\bibitem[{Nolan} {et~al.}(2012)]{Nolan2012}
{Nolan}, P.~L., {Abdo}, A.~A., {Ackermann}, M., {et~al.}, 2012, ApJS, 199, 31

\bibitem[Porth \& Komissarov(2014)]{Porth2014}
Porth, O., \& Komissarov, S.~S.\ 2014, arXiv:1408.3318

\bibitem[{Poutanen} \& {Stern}(2010)]{Poutanen2010}
{Poutanen}, J., \& {Stern}, B., 2010, ApJ, 717, L118

\bibitem[\protect\citeauthoryear{{Rolke}, {L{\'o}pez} \& {Conrad}}{{Rolke} et~al.}{2005}]{Rolke2005}
{Rolke} W.~A.,  {L{\'o}pez} A.~M.,    {Conrad} J.,  2005, Nuclear Instruments
  and Methods in Physics Research A, 551, 493

\bibitem[Saito et~al.(2013)]{Saito2013}
Saito, S., Stawarz, {\L}., Tanaka, Y.~T., et~al.\ 2013, ApJ, 766, L11

\bibitem[{Sikora} {et~al.}(2009)]{Sikora2009}
{Sikora}, M., {Stawarz}, \L., {Moderski}, R., {Nalewajko}, K., \& {Madejski}, G.~M., 2009, ApJ, 704, 38

\bibitem[Sironi \& Spitkovsky(2014)]{Sironi2014}
Sironi, L., \& Spitkovsky, A.\ 2014, \apjl, 783, L21

\bibitem[{Spitkovsky}(2008)]{Spitkovsky2008}
{Spitkovsky}, A., 2008, ApJ, 682, L5

\bibitem[Stern \& Poutanen(2011)]{Stern2011}
Stern, B.~E., \& Poutanen, J.\ 2011, MNRAS, 417, L11

\bibitem[Stern \& Poutanen(2014)]{Stern2014}
Stern, B.~E., \& Poutanen, J.\ 2014, \apj, 794, 8

\bibitem[{Tanaka} {et~al.}(2011)]{Tanaka2011}
{Tanaka}, Y.~T., {et~al.}, 2011, ApJ, 733, 19

\bibitem[Tavecchio \& Ghisellini(2008)]{Tave08}
Tavecchio, F., \& Ghisellini, G.\ 2008, MNRAS, 386, 945

\bibitem[Werner et al.(2014)]{Werner14}
Werner, G.~R., Uzdensky, D.~A., Cerutti, B., Nalewajko, K., \& Begelman, M.~C.\ 2014, arXiv:1409.8262

\end{thebibliography}
\end{document}